\documentclass[conference]{IEEEtran}
\IEEEoverridecommandlockouts
\usepackage{cite}
\usepackage{amsmath,amssymb,amsfonts}
\usepackage{algorithm,algorithmic}
\usepackage{caption}
\usepackage{graphicx}
\usepackage{xcolor}
\usepackage{subfigure}
\usepackage{hyperref}
\usepackage{textcomp}
\usepackage{xcolor}
\newlength\myindent
\newcommand\bindent{%
  \begingroup
  \setlength{\itemindent}{\myindent}
  \addtolength{\algorithmicindent}{\myindent}
}
\newcommand\eindent{\endgroup}

\def\BibTeX{{\rm B\kern-.05em{\sc i\kern-.025em b}\kern-.08em
    T\kern-.1667em\lower.7ex\hbox{E}\kern-.125emX}}

\begin{document}

\IEEEoverridecommandlockouts
\IEEEpubid{\makebox[\columnwidth]{978-1-7281-0858-2/19/\$31.00 \copyright 2019 IEEE \hfill} \hspace{\columnsep}\makebox[\columnwidth]{ }}

\title{SWAG: Item Recommendations using Convolutions on Weighted Graphs
}

\author{\IEEEauthorblockN{}
\IEEEauthorblockA{
Amit Pande, Kai Ni and Venkataramani Kini\\
Data Sciences, Target Corporation
}
}

\maketitle

\begin{abstract}
Recent advancements in deep neural networks for graph-structured
data have led to state-of-the-art performance on recommender
system benchmarks. In this work, we present a 
Graph Convolutional Network (GCN) algorithm SWAG (Sample Weight and AGgregate),
which combines efficient random walks and graph convolutions on weighted graphs
to generate embeddings for nodes (items) that incorporate both
graph structure as well as node feature information such as item-descriptions and item-images. 
The three important \textit{SWAG} operations that enable us to efficiently generate node embeddings based on graph structures are (a) \textit{S}ampling of graph to homogeneous structure, (b) \textit{W}eighting the sampling, walks and convolution operations, and (c) using \textit{AG}gregation functions for generating convolutions. The work is an adaptation of graphSAGE over weighted graphs. We deploy SWAG at Target and train it on a graph of more than 500K products sold online with over 50M edges. 
Offline and online evaluations reveal the benefit of using a graph-based approach and the benefits of weighing to produce high quality embeddings and product recommendations.\end{abstract}

   \section{Introduction}
   
   Convolutional Neural Networks (CNNs) are used to establish state-of-the-art performance on many Computer
Vision applications~\cite{alom2018history}. CNNs consist of a series of parameterized convolutional layers operating locally (around neighboring pixels of an image) to obtain hierarchy of features about an image.  The
first layer learns simple edge-oriented detectors. Higher layers build up on the learning of lower layers to learn more complex features and objects.
The success of CNNs in Computer Vision has inspired efforts to extend the convolutional operation from regular grids (2D images), to graph-structured data~\cite{goyal2018graph}. Graphs, such as social networks, word co-occurrence networks, guest purchasing behavior, protein-protein interactions and communication networks, occur naturally in various real-world
applications. Analyzing them yields insights into the structure of society, language, and different patterns of communication.
In such graphs, a node’s neighborhood is variable sized (each node can have any number of connections to other nodes unlike a pixel which has 8 nearest neighbors and 16 second degree neighbors and that too with a sense of directionality). Generalizing Convolution to graph structures should allow models to learn location-invariant
features. 

The early extension of convolution to graph-structured data~\cite{bruna2013spectral} is  theoretically motivated but not scalable to large graphs as it incurs quadratic computational complexity in number of nodes. Moreover, it requires the graph to be completely observed during training (transductive scenario). Defferrard et al. ~\cite{defferrard2016convolutional}, Kipf \& Welling~\cite{kipf2016semi,kipf2016variational} propose approximations to Graph Convolutions
that are computationally-efficient (linear complexity, in the number of edges). 

Hamilton and Ying~\cite{hamilton2017inductive,ying2018graph} extend graph convolution networks to scenarios where the entire graph is not required during training. In other words, the model learns a function over inputs such as node attributes and node-neighborhood that can be applied to any input graph or node in general, making it more suitable for inductive settings. For example - for a retailer like Target, assortments are frequently updated and thousands of items as well as millions of guests are added every few days. It is desirable to train a model once and let it inductively generate powerful embedding on newer nodes (items or guests) without retraining on the entire dataset.  The high-dimensional information about a node’s neighborhood (graph structure) as well as the node attributes (other higher dimensional information about a graph) can be efficiently condensed or encoded in the form of graph embeddings using unsupervised graph embedding methods for dimensionality reduction. Such embeddings have demonstrated great performance on a number of tasks including node
classification~\cite{hamilton2017inductive,ying2018graph,perozzi2014deepwalk,he2016identity}, knowledge-base completion~\cite{ng2002spectral}, semi-supervised learning~\cite{tang2015line}, and link
prediction~\cite{arora2016simple}. These
node embeddings can then be fed to downstream machine learning systems and aid in tasks such as
node classification, clustering, and link prediction. As introduced by Perozzi et al.~\cite{perozzi2014deepwalk} and Hamilton~\cite{hamilton2017inductive}, these methods operate in two discrete
steps: First, they sample pair-wise relationships from the graph through random walks. Second, they train an aggregation function or an embedding model to learn representations that encode pairwise node similarities.

Recent works have focused on creating an inductive framework for generating node embeddings by leveraging node features (e.g. text attributes, node profile information, node degrees) in order to learn an embedding function from the graph which can be generalized to unknown graphs/nodes. GraphSAGE uses (a) sampling and (b) aggregation operations to  generate higher-quality recommendations than
comparable deep learning and graph-based alternatives at Pinterest~\cite{hamilton2017inductive,ying2018graph}.

However, the links between nodes of a graph convey specific information which is not properly captured by existing architectures. The weights between nodes may signify the cost or advantages or popularity of a transition from one node to another. For example - weights between two nodes in a graph, with each product being a node may represent the probability of co-views, co-purchases, rate of substitution or cost of substitution, depending on the application usage. Traditional product recommendation algorithms such as collaborative filtering~\cite{su2009survey} use this information to deliver product recommendation. In this work, we incorporate weights into the graph based algorithm. The resulting algorithm has three components - (a) \textbf{S}ampling, (b) \textbf{W}eighting and (c) \textbf{AG}gregation  has been abbreviated as\textbf{ GraphSWAG} or simply \textbf{SWAG} in the paper.

The main contributions of this work are as follows:
\begin{enumerate}
\item In this work, we tune graph sampling and aggregation operations by incorporating the knowledge of edge weights into the procedure. Weights in graph are used for sampling, aggregation as well as generation of random walks and measuring loss. 
\item The proposed framework (SWAG) is used for similar or related product recommendations for a  retailer to combine the insights from (a) product or item description (text), (b) item images and (c) purchase behavior (views / add-to-cart / purchases) into a single framework. 
\item The offline as well as online experiments illustrate that such a scheme outperforms image or item attributes based deep learning and unweighted graph based approaches.
\end{enumerate}

This paper is organized as follows. Section II gives an overview of related works. Section III explains the proposed method and the inputs. Section IV gives an overview of the algorithm followed up by experimental results in Section V. Section VI gives conclusions and directions for future work. 
\section{Background \& Related Work}
Our work builds upon recent advances in the field of Graph neural networks (GNNs) or Graph Convolution Networks (GCNs). GCNs are connectionist models that capture the dependence of graphs via message passing between the nodes of graphs. Unlike standard neural networks, graph neural networks retain a state that can represent information from its neighborhood with arbitrary depth. The concept of neural network for graphs was first introduced in ~\cite{chung1997spectral}. Initial  approaches were difficult to train for a fixed point, recent advances in network architectures, optimization techniques, and parallel computation have addressed computational speed issues. The GCNs borrow the image of image convolutions (with small filters) to allow message passing along local neighbors of a node and significantly speed up the model training and convergence. The following properties of graphs are helpful to train GCNs for complex data science tasks. 1) Graphs are the most
typical locally connected structures. 2) The shared weights of GCNs reduce the computational cost compared with traditional spectral
graph theory. 3) multi-layer structure of GCNs  allows us to deal with hierarchical patterns, which captures the features of various sizes. Bruna et al.~\cite{bruna2013spectral} developed an initial GCN based on spectral graph theory. 
 Following on this work, a number of
authors proposed improvements, extensions, and approximations
of these spectral convolutions~\cite{monti2017geometric,zitnik2018modeling,grover2016node2vec,atwood2016diffusion}, leading to new state-of-the-art results on benchmarks such as node
classification, link prediction, as well as web scale recommendations
(e.g., the MovieLens benchmark~\cite{monti2017geometric,hamilton2017inductive,ying2018graph}). These approaches have consistently outperformed techniques based upon matrix factorization
or random walks. Hamilton et al. ~\cite{hamilton2017representation}, Bronstein et al.
~\cite{bronstein2017geometric} and Zhou et al.~\cite{zhou2018graph} provide comprehensive surveys of recent advancements.

The inductive approaches such as GraphSAGE and Pin-SAGE~\cite{hamilton2017inductive,ying2018graph} derive embeddings as a function of node features and neighbors so that the function is scalable or usable over unseen graphs. Instead of training a distinct embedding vector for each node, a set of aggregator functions
 is trained. Each aggregator function aggregates information from a different number of hops, or search depth, away from a given node. 
 The approach presented in this work is an improvement over this work by leveraging graph weights for sampling, aggregation and in unsupervised loss. 
We use a unsupervised loss function to generate the recommendations / embeddings for millions of online items. 
It is a highly scalable GCN framework (can operate on billions of nodes) and based on running local convolutions or aggregations on nodes. For training the model, nodes are selected for the loss function using random walks and negative sampling is used to select negative examples. Using random walks alleviates the requirement of entire adjacency matrix of graph to live in memory.

To our best knowledge, weighted graphs have received little attention. The closest of graphical convolutional neural network (GCNN) with edge information are in G2S~\cite{zitnik2018modeling} and r-GCN~\cite{schlichtkrull2018modeling} for natural language processing. The former uses the edge weights to aggregate the information from neighbors through element-wise multiplication for the state of nodes (See G2S~\cite{zitnik2018modeling} and also equation (8) of \cite{zhou2018graph} for details). However, the edge weights in G2S are learned from the node embeddings through Gates (like GRU). To the contrary, our edge weights are the input to the model. This allows us to incorporate edge information from other sources such as user-browse behavior in graphs. In the latter (r-GCN), the edge weights are used in regularization only because the latter focuses on link prediction, in which regularization plays an important role. 


\section{Proposed Method}

In this section, we describe the technical details of the SWAG algorithm and its implementation for product 
recommendations. 
The key computational blocks of the algorithm is the notion of localized graph convolutions. To generate the embeddings for a node (item), multiple convolutional modules or aggregators aggregate feature information (item descriptions or visual appearance) from the node and its local graph neighborhood. This approach was first proposed in \cite{hamilton2017inductive}. However, all the neighbors are equally treated in this approach. \cite{ying2018graph} proposed using importance sampling to find important neighbors of a node but all important nodes are equally treated. The aggregators use the weights to mix neighbors accordingly. 

\subsection{Problem Setup}
Target is one of the largest general merchandise retailers in US, with Target.com consistently being ranked as one of the most-visited retail Web sites. The website serves millions of product recommendations daily to the guests.

Our task is to generate high quality embeddings of items that can be used for nearest neighborhood lookups and subsequent usage in recommendations. In order to learn these embeddings, we model the shopping behavior of guests at Target as a graph with each node representing an item. In addition to the graph structure, we assume that the items are associated with additional features i.e. metadata or content information about the item. Each item is associated with rich item descriptions and image features. The learnt embeddings are to be used for product recommendations.

\subsection{Generating graph weights}
In this work, we try two methods to generate graph weights from behavioral observations of aggregate guest shopping behavior.
\subsubsection{Jaccard Index}
The edges of graph are weighted according to past customer views. Therefore our graph has weights on all its edges. The weights of the graph are generated by the Jaccard Index. To be specific, we calculate the relative frequency of views for each pair of items and then make an 
$\arctan$ transformation of that relative frequency. 
For online items $i$ and $j$, the relative frequency $F(i, j)$ is defined as follows:
\begin{equation}\label{relative frequency}
F(i, j) = \frac{VC(i \cap  j)}{VC(i\cup j)}
\end{equation}
where $VC(i\cap  j)$ is the view counts or the number of guests that view item $i$ and $j$ in one 
session and $VC(i\cup  j)$ is the view counts for either item $i$ or $j$ being viewed in a 
session. In online retail, the relative view frequency $F(i, j)$ for item $i$ and $j$ is usually very 
small. $3\%$ common view is already a very big number for a pair of items, we divide the relative 
frequency by the median of frequency in one category to scale into the weight function $s: 
(\mathcal{V}, \mathcal{V})\rightarrow(0, 1)$, which is defined to be
\begin{equation}\label{weightofedge}
s(i, j) = \frac{2}{\pi}\times\arctan(\frac{F(i, j)}{\text{median of } F}).
\end{equation}
After the transformation, the weights are closer to uniform distribution on $[0, 1]$ across the edges. 


\subsubsection{Weighted co-occurrences}:
In another approach, we generate weights not just using co-view counts but also give weights to add-to-cart and ultimately bought together counts. 
The different activities of guests, such as view / add to cart / purchase of products are weighed using empirically determined weights. Further, we apply time decay on co-occurrences to capture the recency of items. 
The weighted co-occurrence of two products $i$ and $j$ for $N$ customer sessions is given by: 
\begin{equation}
s(i,j) = \sum _ {s=1} ^{s=N}  W(i)  W(j) / Rec(s)
\end{equation}
where, $ W(i) $ and $ W(j) $ are highest weights of products $i$ and $j$ in session $s$, $Rec(s)$ is recency of session $s$ (if session has occurred $10$ days back then $Rec(s)$ is $10$).

Finally, we normalize the weights per node and apply arctan transform to set the weights in range $[0,1]$.

\subsection{Generating node embeddings}

The image embeddings are generated using the pre-trained VGG-16 model~\cite{simonyan2014very}. The last fully connected layers are not used and we use the output up to convolutional layers and max-pool layers (the last layer is used as average-pool instead of max-pool). 

The item embeddings are obtained by training a word embedding model~\cite{mikolov2013distributed} on item attributes and descriptions in our item catalog.


\section{Algorithm}
In this section, we describe the technical details of the SWAG architecture and training, as well as a GPU pipeline, to efficiently generate embeddings using a trained SWAG model. The model has two main steps - sampling and aggregation. We introduce the notation used in the paper in Table \ref{noteswag}.

\begin{table}[]
\begin{tabular}{|l|l|}
\hline
Notation                          & Explanation                                                                                                                                          \\ \hline
$\mathcal{G}$                     & the given graph
 \\ \hline
$\mathcal{V}$                     & the node set of $\mathcal{G}$                                                                                                                        \\ \hline
$\mathcal{E}$                     & the edge set of $\mathcal{G}$                                                                                                                        \\ \hline
$u\in\mathcal{V}$                 & a certain node $u$ of the graph                                                                                                                      \\ \hline
$(u, v)\in\mathcal{E}$            & a certain edge of the graph                                                                                                                          \\ \hline
$s(u, v)$                         & the weight of edge $(u, v)\in\mathcal{E}$                                                                                                            \\ \hline
$r(u, v)$                         & \begin{tabular}[c]{@{}l@{}}the maximum geometric mean of weights along paths\\ connecting $u$ and $v$; $u$ and $v$ may not be neighbors\end{tabular} \\ \hline
$\mathcal{N}$ or $\mathcal{N}(u)$ & the neighboring function on $\mathcal{G}$ or the neighbors of node $u$                                                                               \\ \hline
$h_u^k$                           & the hidden state of node $u$ at $k$-th layer                                                                                                         \\ \hline
$\mathbb{W}^k$                    & the linear parameter of neural network at depth $k$                                                                                                  \\ \hline
$\sigma$                          & \begin{tabular}[c]{@{}l@{}}the non-linearity function for network, use ReLU for all\\  layers except the last one.\end{tabular}                      \\ \hline
$\pi_k$                           & \begin{tabular}[c]{@{}l@{}}the aggregator function applied to neighbors at depth k. \\ Choices are GCN, mean aggregator, LSTM etc.\end{tabular}      \\ \hline
$\alpha$                          & the positive degree on path weights inside loss function                                                                                             \\ \hline
$\beta$                           & the positive degree on edge weights inside sampling                                                                                                  \\ \hline
$\gamma$                          & the positive degree on edge weights inside aggregation                                                                                               \\ \hline
\end{tabular}
\caption{Notation in Swag Algorithm.}
\label{noteswag}
\end{table}

\subsection{Sampling}
Sampling is very important in Graph Convolutional Networks. As opposed to computer vision, where convolutional neural networks can use pixel proximity as a feature, GCNs do propagation guided by the graph structure \cite{zhou2018graph}. Therefore, for any given node, we need to efficiently select its neighbors for convolution. In Swag Algorithm, the neighbor function, 
\begin{equation}\label{eq-sampling}
\mathcal{N}_s: \mathcal{V} \rightarrow 2^{\mathcal{V}}
\end{equation}
samples a subset of neighbors for any given node $v\in\mathcal{V}$ based on the edge weights of 
its neighbors. In contrast to  prior work\cite{hamilton2017inductive}, in which the neighbor function 
selects neighbors uniform randomly, we select neighbors with probability proportional to $s(u, v)^{\beta}$, where $ s(u, v)$ is the weight of edge $(u, v)\in\mathcal{E}$ and $\beta$ is a sampling degree hyper-parameter in $[0, \infty)$. 
In our use case of product recommendations, the larger the weight of the edge, the more chances that the corresponding neighbor should be selected in sampling.
When $\beta = 0$, the impact of weights is neutralized. On the other hand, larger value of $\beta$ implies that only neighbors with large weights will get selected. We formalize the sampling algorithm as follows. Each layer of SWAG can have distinct number of sampled neighbors, so the algorithm below will be applied to each layer of the neural network.

\begin{algorithm}[htb]
\caption{Sampling: SWAG embedding generation }
\begin{algorithmic}
    \STATE \textbf{Input:} Graph $\mathcal{G}(\mathcal{V}, \mathcal{E})$ and a weight function $s(u, v)$ for any $(u, v)\in\mathcal{E}$, a sampling hyper-parameter $\beta$
    \STATE \textbf{Output:} Graph with homogeneous number of neighbors.
    \FORALL{$u \in \mathcal{V}$}
        \bindent
       \STATE $\omega(u,v)$ = $k s(u, v)^\beta$, s.t.$ \sum_{v\in\mathcal{N}(u)} \omega(u,v) =1$;
       \STATE and sample $v\in\mathcal{V}$ based on $\omega(u,v)$ 
         \eindent
      \ENDFOR
\end{algorithmic}
\end{algorithm}

\subsection{Aggregation}
After sampling, the selected neighbors need to be aggregated to their corresponding nodes for information clustering. 
The aggregation step is similar to convolution over nearby pixels in images and has the goal of aggregating information from neighboring nodes. However, a node's neighbors have no particular or natural ordering in graphs. The \textit{mean} aggregator, for example, would take a element-wise weighted mean of vectors in $w^{\gamma}\mathrm{h}_u^{k-1}, \forall u \in \mathcal{N}(v)$. The max-pool operator would take the max of the weighted embeddings and so on. 
We formulate the aggregation algorithm as follows.

\begin{algorithm}[htb]
\caption{Aggregation: SWAG embedding generation}
\begin{algorithmic}
    \STATE \textbf{Input:} Graph $\mathcal{G'}(\mathcal{V'}, \mathcal{E'})$: input features $\{\mathrm{x}_{v}, \forall v\in\mathcal{V}\}$; depth $K$; weight matrices $\mathbb{W}^k, \forall k \in \{1, \ldots, K\}$; non-linearity $\sigma$; differentiable aggregator functions $\pi_k, \forall k \in \{1, \ldots, K\}$; neighborhood function $\mathcal{N}: \mathcal{V} \rightarrow 2^{\mathcal{V}}$; edge weight function $s(u, v), \forall(u, v)\in\mathcal{E}$.
    \STATE \textbf{Output:} Vector representations $\mathrm{z}_v$ for all $v \in \mathcal{V}$
    \STATE $\mathrm{h}_v^{0}\leftarrow x_v / \|x_v\| , \forall v\in\mathcal{V}$.
    \FORALL{$k \in \{1, \ldots, K\}$}
        \bindent
        \FORALL{$v \in \mathcal{V}$}
            \STATE  \vspace{-0.1in} \hspace{-0.3in}
            \begin{equation}\label{eq:agg-neigh}
            \mathrm{h}_{\mathcal{N}(v)}^k \leftarrow\pi_k(\{s(u, v)^{\gamma}\mathrm{h}_u^{k-1}, \forall u \in \mathcal{N}(v)\})
            \vspace{-0.1in}
            \end{equation}
            \STATE \vspace{-0.1in}
            \begin{equation*}
            \mathrm{h}_{\mathcal{N}(v)}^k \leftarrow \mathrm{h}_{\mathcal{N}(v)}^k / \|\mathrm{h}_{\mathcal{N}(v)}^k\|
            \end{equation*}
            \STATE \vspace{-0.3in}
           \begin{equation}\label{eq:concat}
           \mathrm{h}_v^k \leftarrow \sigma(\mathbb{W}^k \cdot \textnormal{CONCAT}(\mathrm{h}_v^{k-1}, \mathrm{h}_{\mathcal{N}(v)^k}))
           \end{equation}
         \ENDFOR
         \eindent
         \STATE 
        $  \mathrm{h}^k_v \leftarrow \mathrm{h}^k_v / \|\mathrm{h}^k_v\|$
      \ENDFOR
      \STATE 
    $     \mathrm{z}_v \leftarrow \mathrm{h}_v^K, \forall v\in\mathcal{V}$
\end{algorithmic}
\end{algorithm}

In aggregation, if there are two sources of input features $(\mathrm{x}^1_v, \mathrm{x}^2_v)$ as in our case embeddings from image and text, we can combine them together as
\begin{equation*}
\mathrm{x}_v = \sigma(\mathrm{x}^1_{v} + \mathbb{W}^0 \cdot \mathrm{x}^2_{v}), \forall v\in\mathcal{V},
\end{equation*}
where $\mathbb{W}^0$ is some linear transformation matrix to make $\mathrm{x}^1_{v}$ and $\mathrm{x}^2_{v}$ in the same dimension and it is a trainable parameter in training; $\sigma$ is a nonlinear element-wise function.

The intuition behind the algorithm is that at each iteration, or search depth, nodes aggregate information from their local neighbors, and as this process iterates, nodes incrementally gain more and more information from further reaches of the graph from their neighbors. Unlike prior works~\cite{hamilton2017inductive}, the hidden state $h_u^{k-1}$ here is discounted using the edge weight in aggregation to the state of node $v$. Our guideline for this multiplicative factor $s(u,v)^{\gamma}$ is to incorporate the importance of item-to-item view dependency so that higher weights are aggregated more than lower ones. The parameter $\gamma\in[0, \infty)$ is neutralized when it is set to zero. For larger values of $\gamma$ the neighbors with higher weights contribute more to the aggregation.

The aggregation function $\pi_k, \forall k\in\{1, \ldots, K\}$ could be one of those in \cite{hamilton2017inductive}: Mean aggregator, LSTM aggregator, Pooling aggregator, node2vec~\cite{grover2016node2vec}, GCN~\cite{kipf2016semi}.


\subsection{Loss function}

The Sampling and Aggregation operations are forward propagation operations, i.e. we are assuming that the weights and hyper-parameters are already learnt. The model parameters can be learned using standard
stochastic gradient descent and back-propagation techniques using the loss function described in this section.
In order to learn useful, predictive representations in a fully unsupervised setting, we apply a graph-based loss function to the output representations, $\mathrm{z}_u, \forall u\in\mathcal{V}$, and train parameters $\mathbb{W}^k$ of the aggregator functions  in equation (\ref{eq:concat}) for $k \in \{1, \ldots, K\}$ via stochastic gradient descent. The graph-based loss function encourages nearby nodes to have similar representations, while enforcing that the representations of disparate nodes are highly distinct:
\begin{eqnarray}
\mathcal{L}_{\mathcal{G}}(\mathrm{z}_u) &=& -r(u, v)^{\alpha}\log(\sigma(\mathrm{z}_u^{\top}\mathrm{z}_v))\nonumber\\ 
&-& Q\cdot\mathbb{E}_{v_n\sim \mathbb{P}_n(v)}\log(\sigma(-\mathrm{z}_u^{\top}\mathrm{z}_{v_n})),\label{eq:loss}
\end{eqnarray}
where $v$ is a node that co-occurs near $u$ on fixed-length random walk, $\sigma$ is the sigmoid function, $\mathbb{P}_n$ is a negative sampling distribution, and Q defines the number of negative samples. $r(u, v)$ is the accumulated mean of the weights on the random walk for node $u$ and $v$ and $\alpha$ is another hyper-parameter to be tuned for the exponential degrees of on weights of random walks. In our implementation, we choose geometric mean of the weights along the random walk for $r(u, v)$. Other ways of combining edge weights include arithmetic and maximum of weights of edges along the path and we will leave the exploration to the future work.

By adding the weights $r(u, v)^{\alpha}$ into the loss function (\ref{eq:loss}), the algorithm will be more focused on minimizing the distance between node $u$ and $v$ with larger edge weights. Since SWAG algorithm tries to get the embeddings with larger relative view frequency items closer, the weighted loss function is more useful to our purpose.


\section{Experiments}
We evaluate the embeddings generated by SWAG to recommend related products to guests when they click on a product and reach product display page. To recommend related products or items, we select the K nearest neighbors to the query item in the embedding space. The performance on this task is evaluated both online and offline. 
\begin{figure*}[htb]%
\centering
\subfigure[Clothing]{%
\label{clothing-beta}%
\includegraphics[width=4.3cm, height=2.5cm]{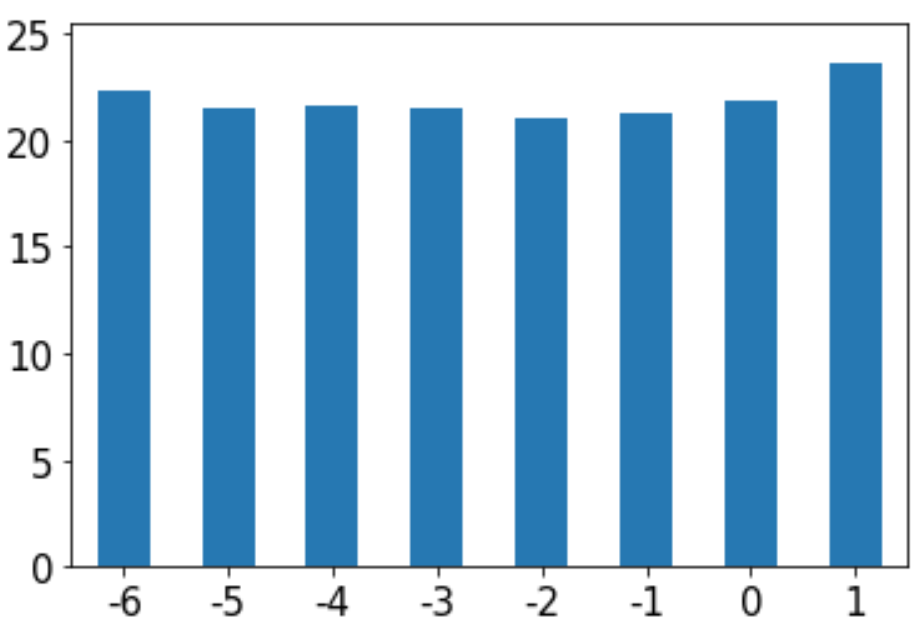}}%
\hspace{3pt}%
\subfigure[Home]{%
\label{home-beta}%
\includegraphics[width=4.3cm, height=2.5cm]{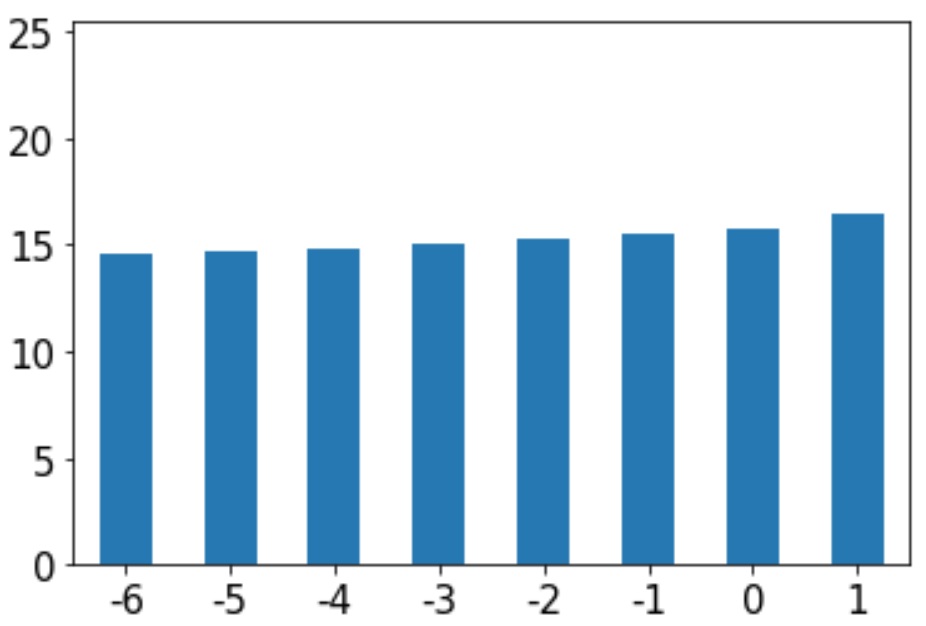}} 
\subfigure[Baby]{%
\label{baby-beta}%
\includegraphics[width=4.3cm, height=2.5cm]{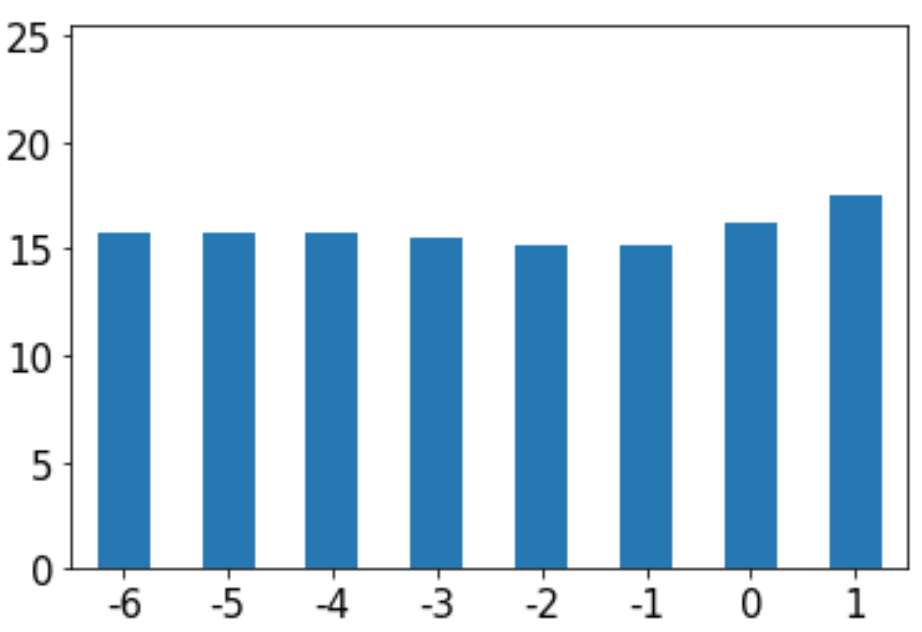}}%
\hspace{3pt}%
\subfigure[Electronics]{%
\label{elec-beta}%
\includegraphics[width=4.3cm, height=2.5cm]{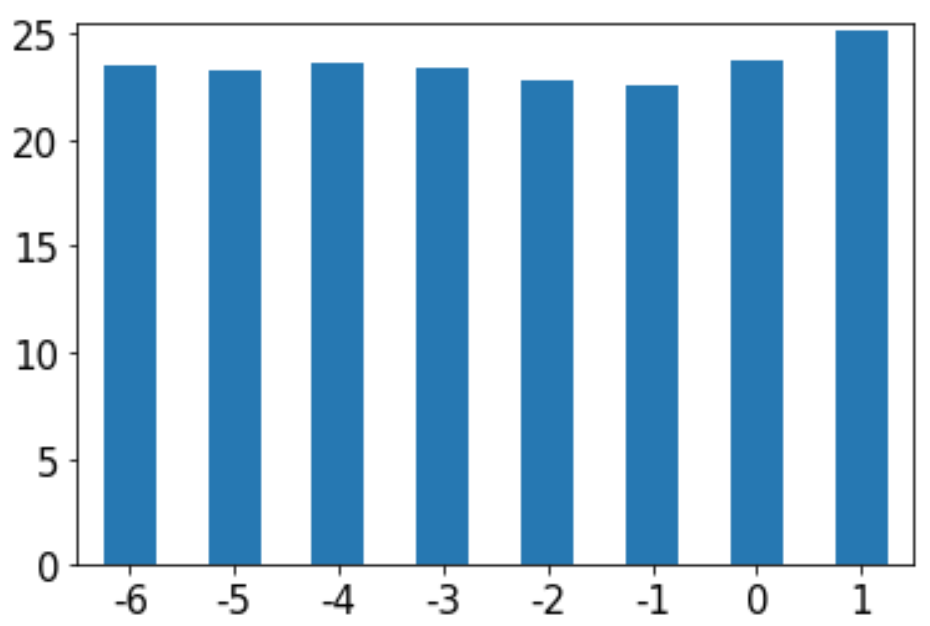}}%
\caption[]{Impact of $\beta$ (sampling hyperparameter) in view rates. The x-axis is the logarithmic value of $\beta$ with base $10$. The best view rates are obtained for $\gamma=10$}
\vspace{-0.05in}
\label{im-beta}%
\end{figure*}

\subsection{Setup}
The loss function of SWAG is unsupervised, hence it implies that it tries to bring neighboring items closer in embedding space and also bring high weighted neighbors closer than low weighted neighbors. In our tasks, we actually train four models for four distinct categories of merchandise: clothing or apparel, baby, home products and electronics items. We choose these four separately because the co-views or co-purchases across one such category are found to be more relevant for the guests than cross-category. Moreover, we train four different models as we assume that the role of item embeddings or image embeddings or past guest behavior would differ depending on the category. Intuitively, image embeddings may play a large role in apparel selection than purchasing an iPhone. The total number of training nodes is close to 500K and graphs have close to 50M edges. The graph is generated using the guest's interaction with the retailer's website (billions of touchpoints).


For offline evaluation, we take past session logs of online guest behavior. We set up an offline evaluation where we evaluate the performance of these embeddings against past guest sessions. For example, if a guest viewed item A and then viewed items B,C,D,E and F  in a past session, we assume A to be the seed item and B/C/D/E/F to be the actual views of the guest. We compare this to the recommendations from the model in consideration and calculate the actual view rate. 
View rate thus defined as the percentage of guests who looked at top N recommendations (N is typically set to 5 as most guest look at top 5 recommendations only) and clicked on one of them. However, this simulation is based on past traces of online behavior and the guests were not actually shown to the recommendations.

We apply word2vec algorithm on item description and item attributes to generate the $200$-dimensional embeddings for online items on Target.com in these four categories. For image embeddings, we tried both the VGG-16 and ResNet-50 models from ImageNet to generate the image embeddings for online items on Target.com. On evaluation, we settled on VGG-16 model as the embeddings performed slightly higher than ResNet-50 for our task and our product catalog.  VGG-16 embeddings are 512 dimension vector while ResNet embeddings were 2048 dimensional vector. The size of input embeddings has an inverse relation with the computational speed of model.
The weights are generated using the logic mentioned above by combining the co-views, add-to-cart and purchase behaviors of the guests on a 200 day window for each item, weighing for recency and normalizing it.

\begin{figure*}[htb]%
\centering
\subfigure[Clothing]{%
\label{clothing-gamma}%
\includegraphics[width=4.3cm, height=2.5cm]{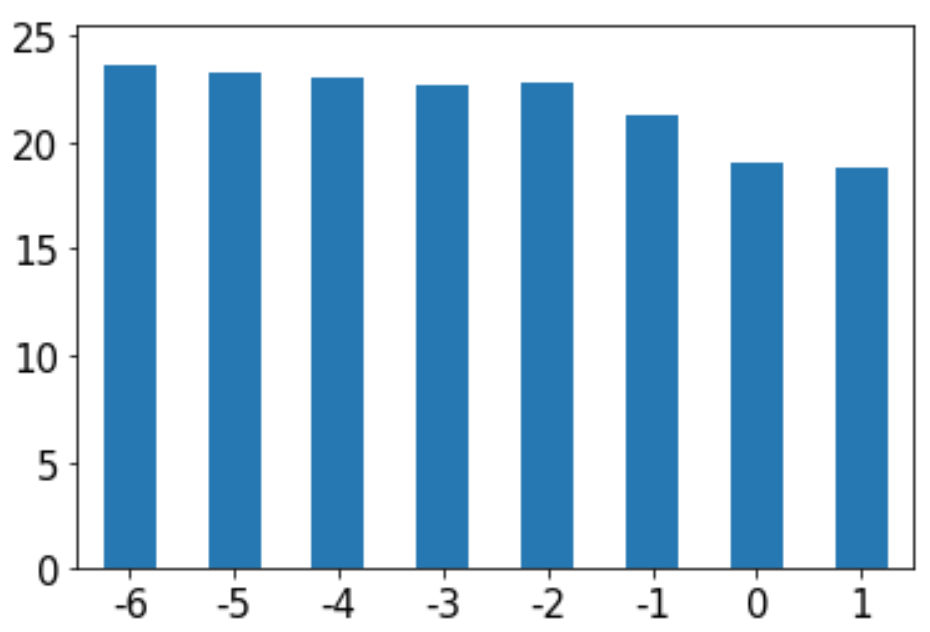}}%
\hspace{3pt}%
\subfigure[Home]{%
\label{home-gamma}%
\includegraphics[width=4.3cm, height=2.5cm]{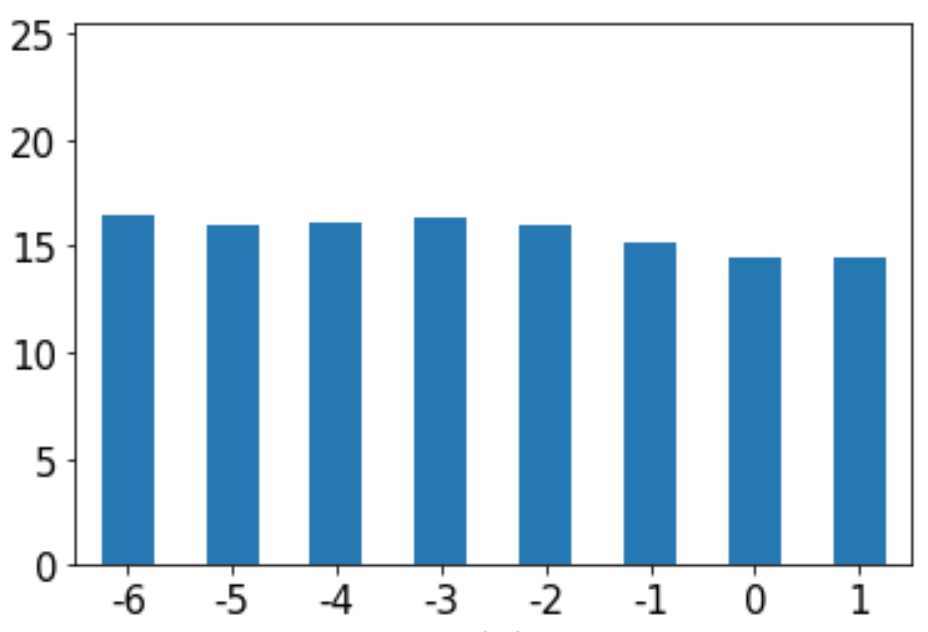}} 
\subfigure[Baby]{%
\label{baby-gamma}%
\includegraphics[width=4.3cm, height=2.5cm]{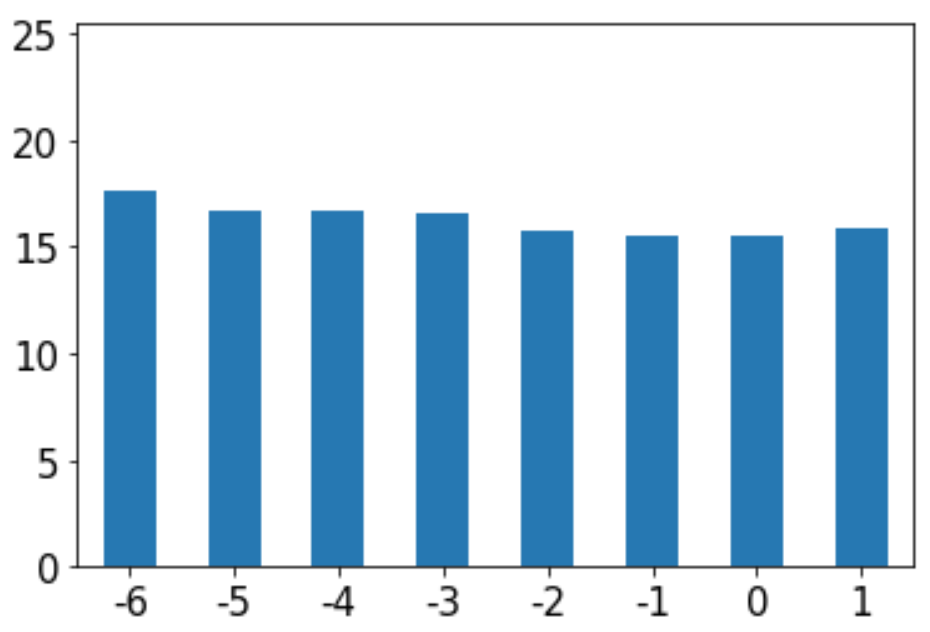}}%
\hspace{3pt}%
\subfigure[Electronics]{%
\label{elec-gamma}%
\includegraphics[width=4.3cm, height=2.5cm]{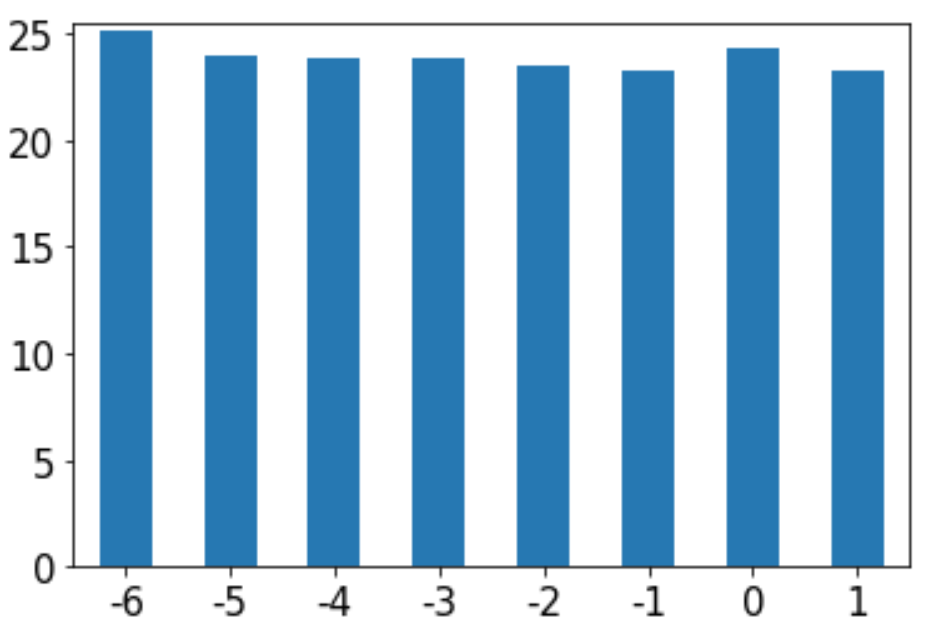}}%
\caption[]{Impact of $\gamma$ (aggregation hyperparameter) in view rate..  The x-axis is the logarithmic value of $\gamma$ with base $10$. The best view rates are obtained with non-zero $\gamma$ in each category.}
\vspace{-0.05in}
\label{im-gamma}%
\end{figure*}

\begin{figure*}[htb]%
\centering
\subfigure[Clothing]{%
\label{clothing-alpha}%
\includegraphics[width=4.3cm, height=2.5cm]{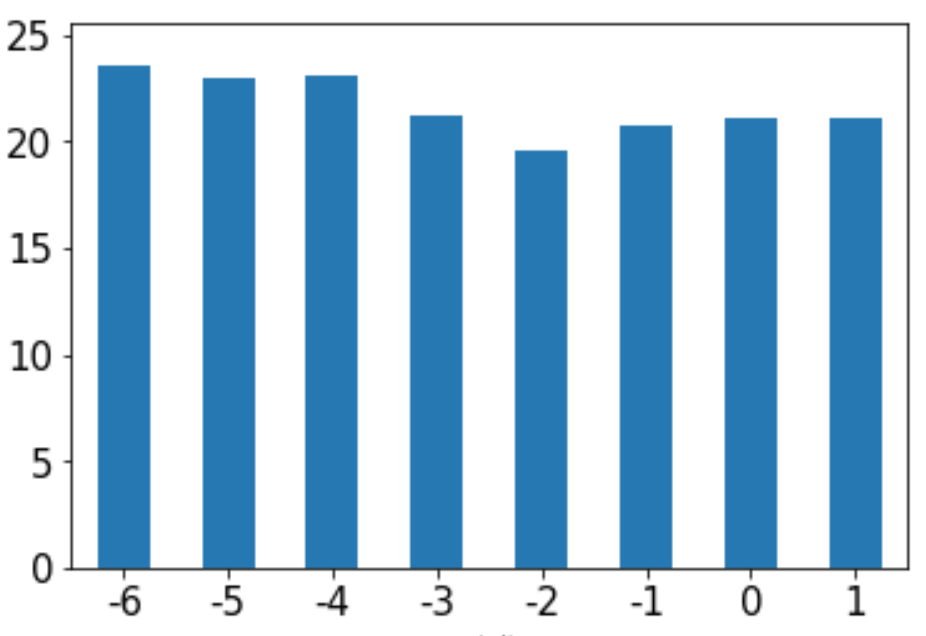}}%
\hspace{3pt}%
\subfigure[Home]{%
\label{home-alpha}%
\includegraphics[width=4.3cm, height=2.5cm]{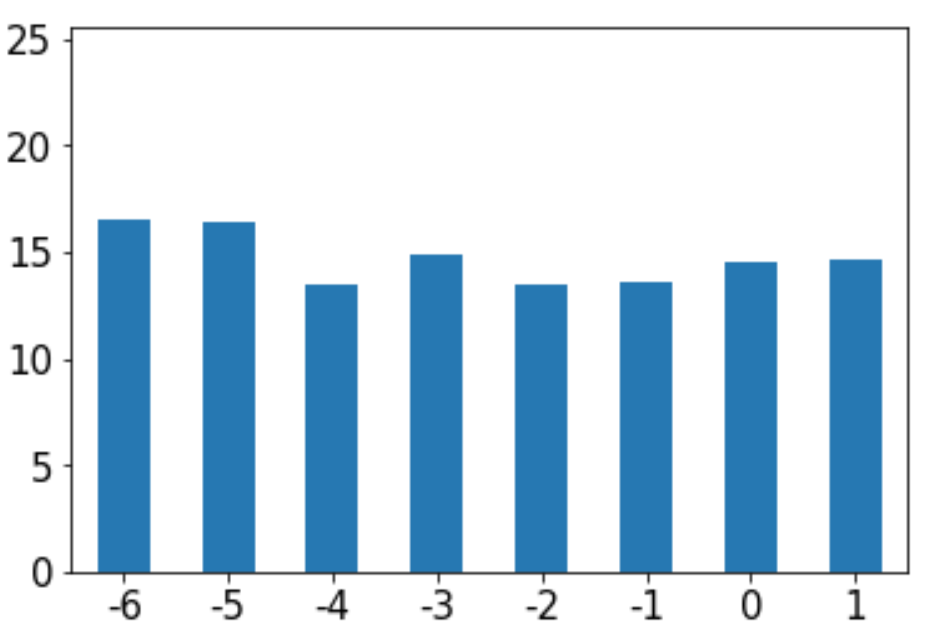}} 
\subfigure[Baby]{%
\label{baby-alpha}%
\includegraphics[width=4.3cm, height=2.5cm]{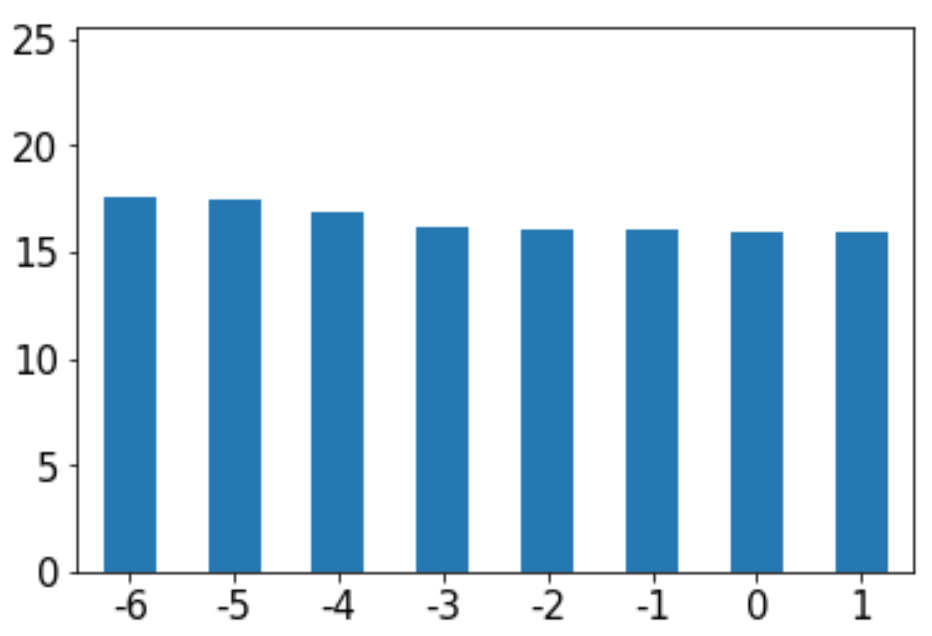}}%
\hspace{3pt}%
\subfigure[Electronics]{%
\label{elec-alpha}%
\includegraphics[width=4.3cm, height=2.5cm]{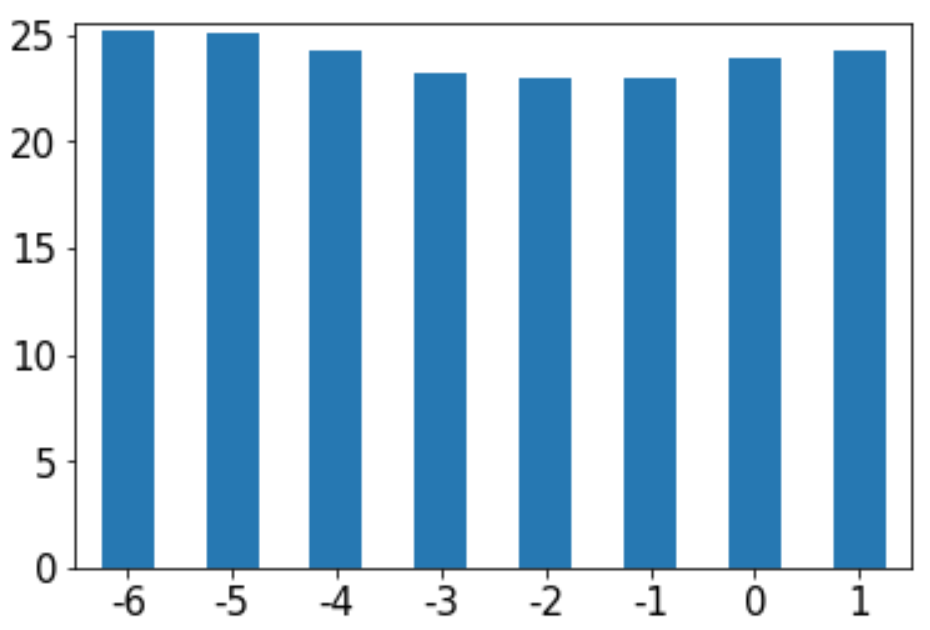}}%
\caption[]{Impact of $\alpha$ (loss hyperparameter) in view rate.  The x-axis is the logarithmic value of $\alpha$ with base $10$. The best performance is reached by non-zero values of $\alpha$ for each category.
}
\vspace{-0.05in}
\label{im-alpha}%
\end{figure*}

\begin{figure}[h!]%
\centering
\subfigure[View rate]{%
\label{view-rate}%
\vspace{-0.2in}
\includegraphics[width=4cm, height=3cm]{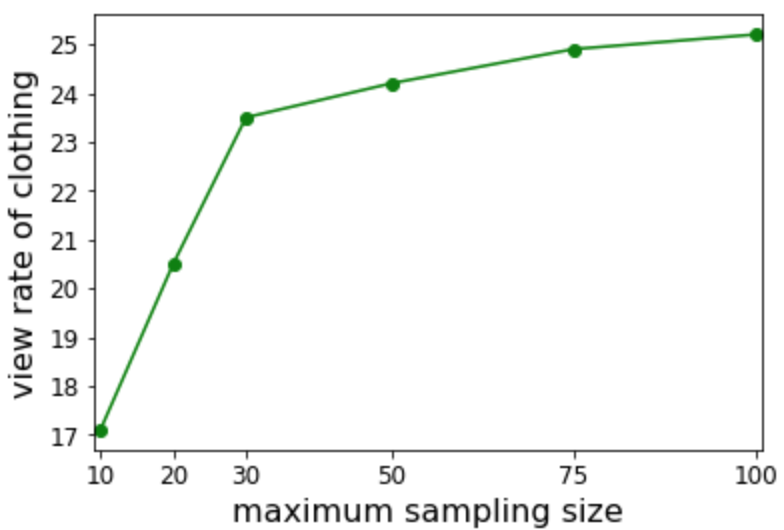}}%
\qquad
\subfigure[Computational time]{%
\label{run-time}%
\includegraphics[width=4cm, height=3cm]{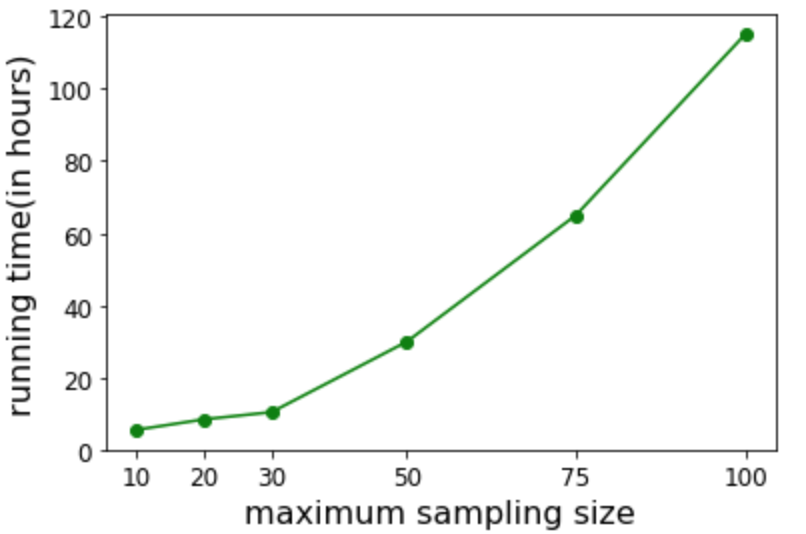}}%
\caption[]{View rate and run time comparisons for different size of samples (neighbors). Computational time is calculated  for two P100 GPU nodes with $200$G memory for clothing category. X-axis represents the size of sample used.}%
\label{vr-time-bench}%
\vspace{-0.1in}
\end{figure}
To get best hyper-parameter sets for each configuration (SWAG for a particular category of items for input being (a) item description, (b) item image, (c) both), we use the skopt package for tuning~\footnote{https://scikit-optimize.github.io/}. Here the loss function is set to maximize the view rate. We tune the hyper-parameters of $\alpha$, $\beta$ and $\gamma$ in the uniform logarithmic range of $[0, 10]$. At the $0$ end, the algorithm is the same as the GraphSage algorithm in \cite{hamilton2017inductive}. At the high $10$ end, the weights raised to power of $10$ have been significantly reduced. We discuss the impacts of hyper-parameters $\alpha$, $\beta$ and $\gamma$ below.

The training was conducted on GPU nodes with (up to) 377 GB of RAM, 40 CPUs (Xeon 2.2GHz) and two P100 GPUs each. 

In our offline evaluation, it is observed that the view rates are higher for clothing and electronics categories and lower for home and baby. These differences are due to market trends and seasonality associated with the past browse events. For example - during the festival season more people shop electronics than baby or home items. They don't conclude anything about the performance of the algorithm per se. 

\subsection{Impacts of weight hyper-parameters $\alpha$,$\beta$ and $\gamma$}
The impact of hyper-parameter $\beta$, which is the sampling exponential degree on edge weights, is shown in Figure \ref{im-beta}. In Figure \ref{im-beta}, we take the average ratio among all of training outputs for $\beta$ in the range of $10^{-6}$ to $10$ and the $x$-axis is the logarithmic of $\beta$.
In all categories, increasing $\beta$ significantly improves the view rate, particularly when $\beta=10$. Please note that the impact of sampling hyper-parameters is illustrated in Figures for SWAG models with item-description embeddings (ID) as the input (for reasons mentioned later). However, the trend was the same for all other inputs.

The view ratios are higher for $\gamma$ closer to zero ($\gamma\approx{10^{-6}}$). In Figure \ref{im-gamma},  we observe significant dip in view rates for clothing as compared to other categories. Typically, clothing or apparel is a category where guests browse the most and across multiple categories before purchasing. Hence, a lot of edges can be spurious or irrelevant (with lower weights). Weighted aggregation seems to improve the performance by lowering the weightage to low weight neighbors.

Figure \ref{im-alpha} plots the impact of loss hyperparameter on view rates across the categories.  The impact of loss degree $\alpha$ is not significant. The bar-plots of the ratio has almost the same height for the choice of $\alpha$'s in all categories. But we can observe that low values of $\alpha$ have a slightly higher view rate across all categories.

\begin{figure*}[htb]%
\centering
\subfigure[Clothing]{%
\label{clothing-agg}%
\includegraphics[width=4.3cm, height=4.5cm]{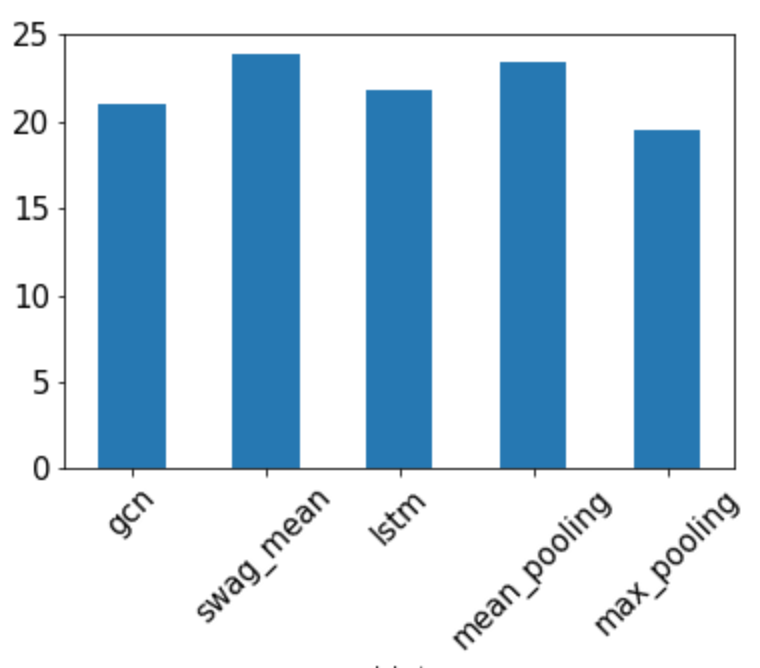}}%
\hspace{3pt}%
\subfigure[Home]{%
\label{home-agg}%
\includegraphics[width=4.3cm, height=4.5cm]{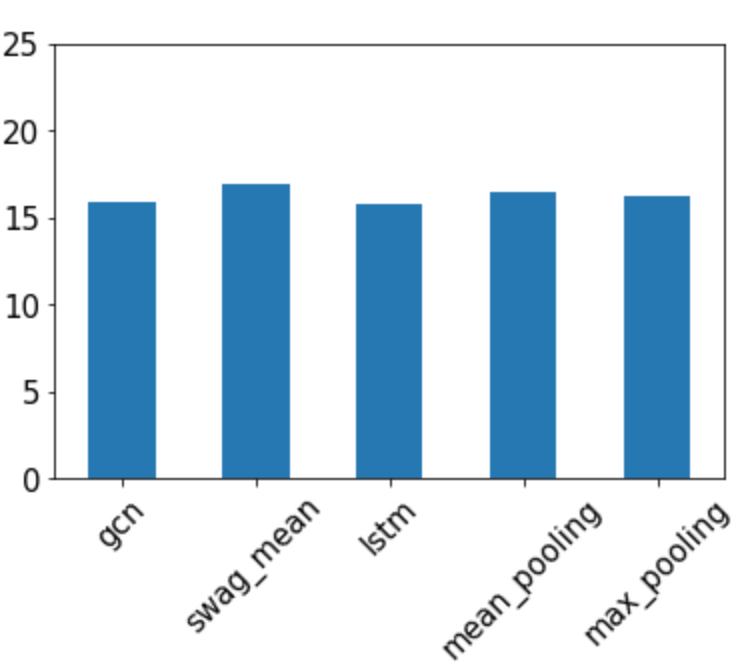}} 
\subfigure[Baby]{%
\label{baby-agg}%
\includegraphics[width=4.3cm, height=4.5cm]{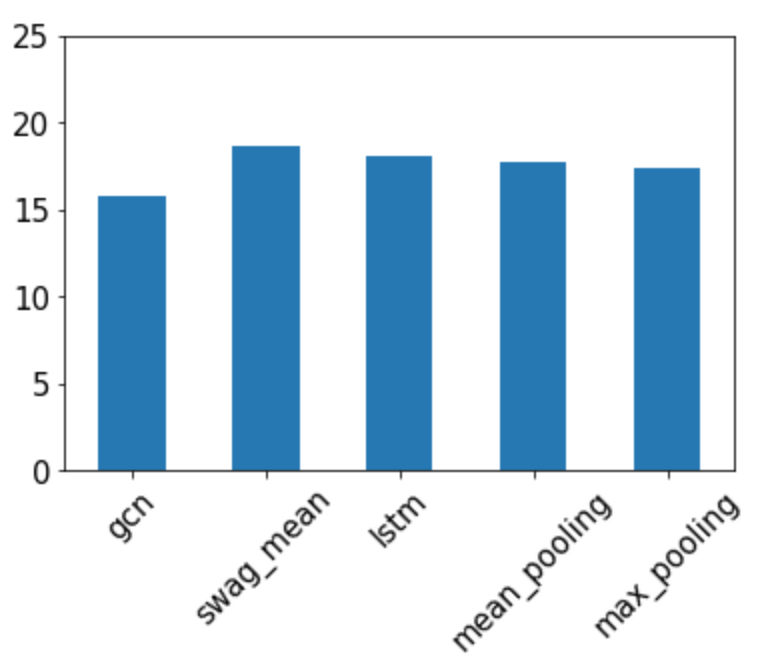}}%
\hspace{3pt}%
\subfigure[Electronics]{%
\label{elec-agg}%
\includegraphics[width=4.3cm, height=4.5cm]{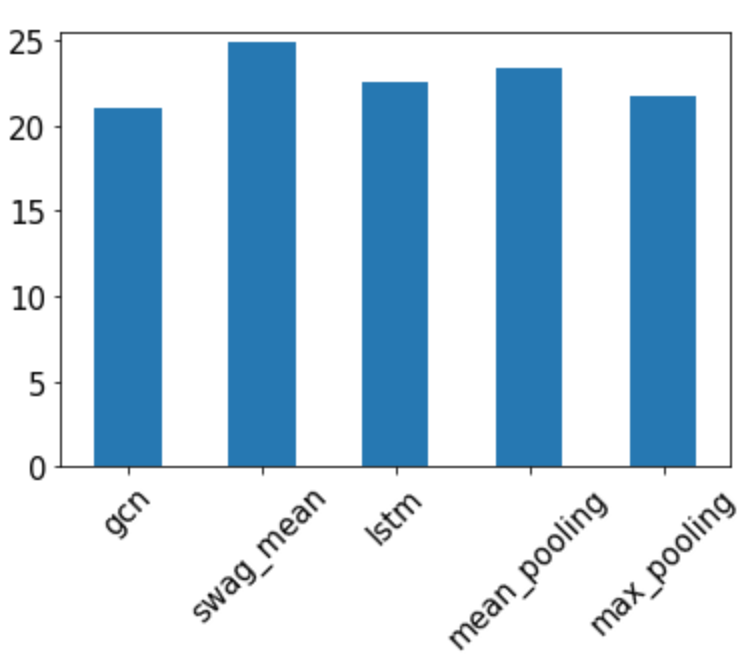}}%
\caption[]{Impacts of different aggregators in training. }
\vspace{+0.1in}
\label{agg-func}%
\end{figure*}

\begin{figure*}[bth]
\centering
\includegraphics[height = 10cm]{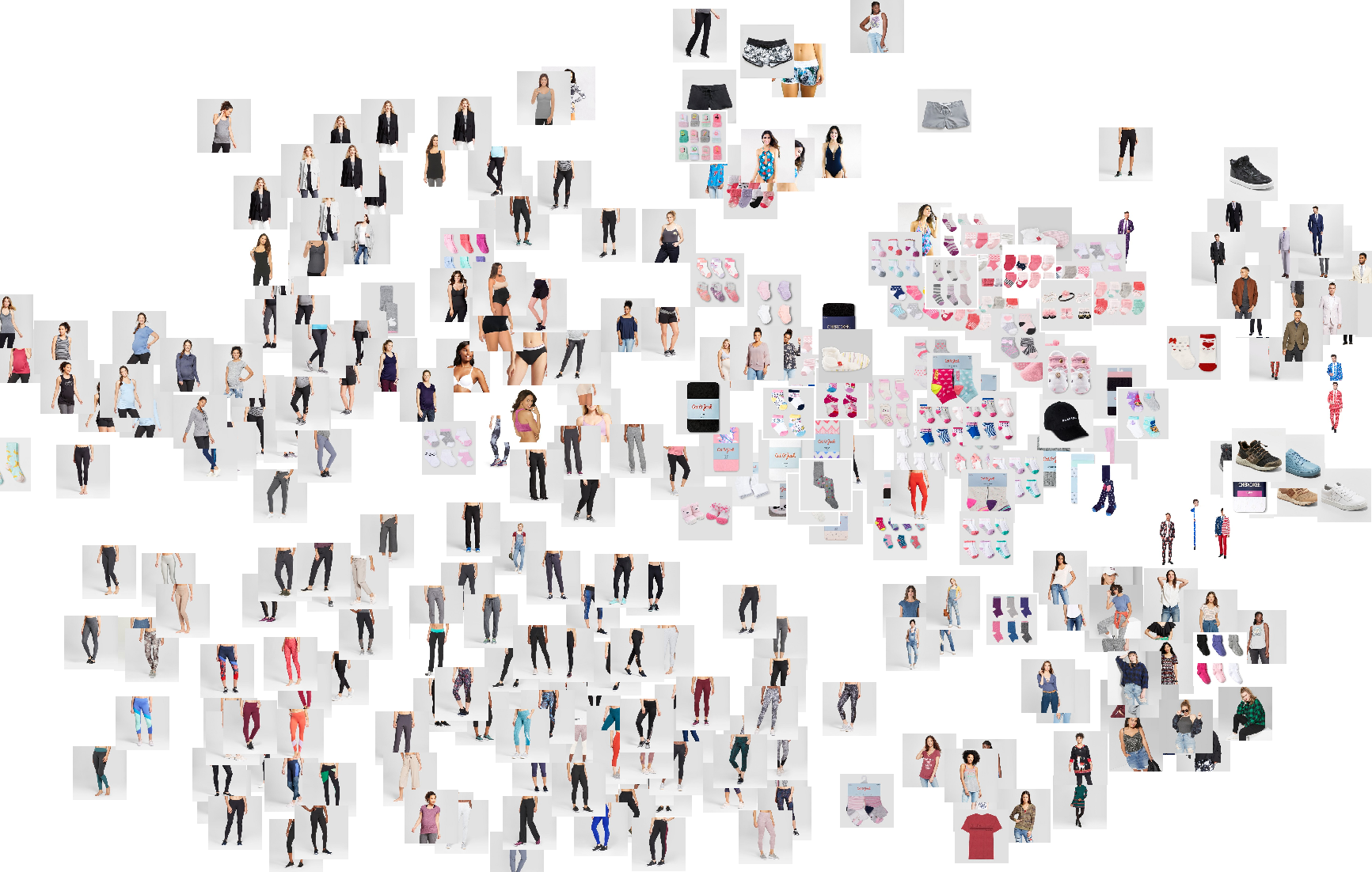}
\caption[Caption for LOF]{t-SNE plot of embeddings for $300$ items in 2 dimensions.}
\label{img-tsne}
\end{figure*}

\begin{figure*}[htb]%
\centering
\subfigure[Clothing]{%
\label{cos_dist_clo}%
\includegraphics[width=4.25cm, height=4cm]{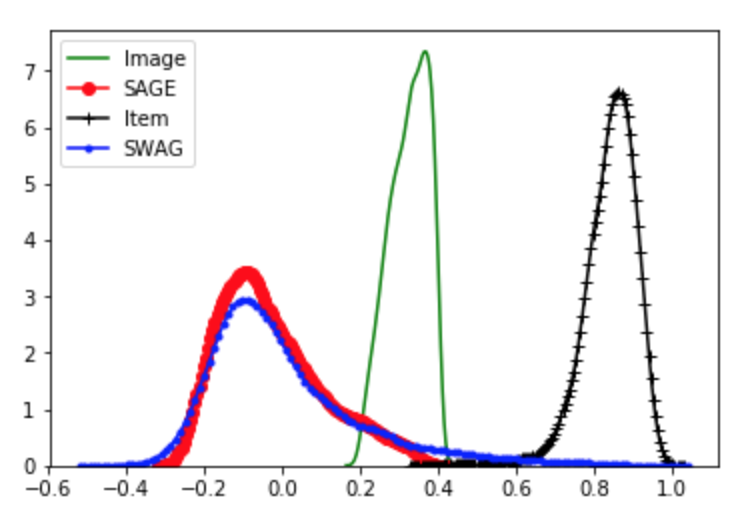}}%
\hspace{3pt}%
\subfigure[Home]{%
\label{cos_dist_home}%
\includegraphics[width=4.25cm, height=4cm]{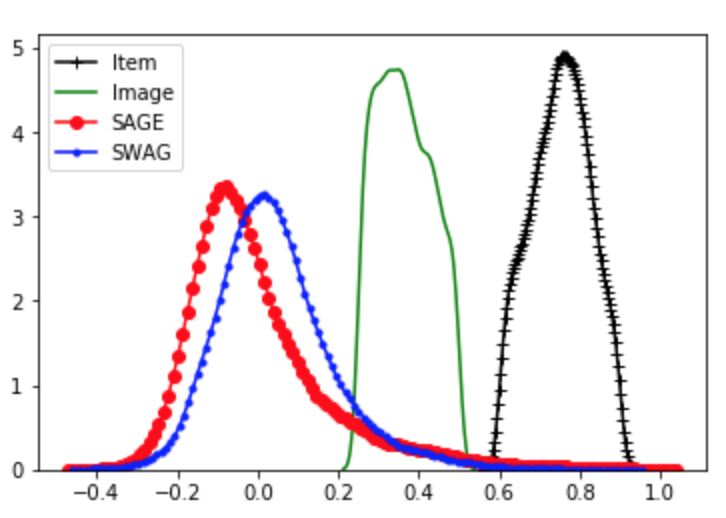}}%
\hspace{3pt}%
\subfigure[Baby]{%
\label{cos_dist_baby}%
\includegraphics[width=4.25cm, height=4cm]{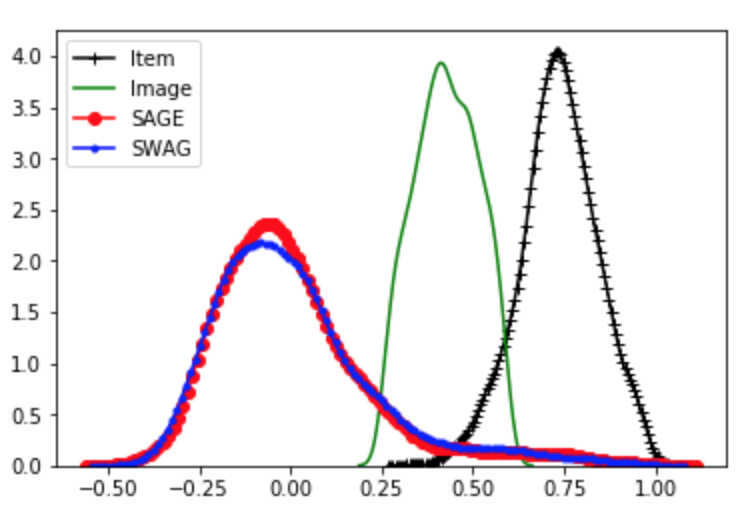}}%
\hspace{3pt}%
\subfigure[Electronics]{%
\label{cos_dist_elec}%
\includegraphics[width=4.25cm, height=4cm]{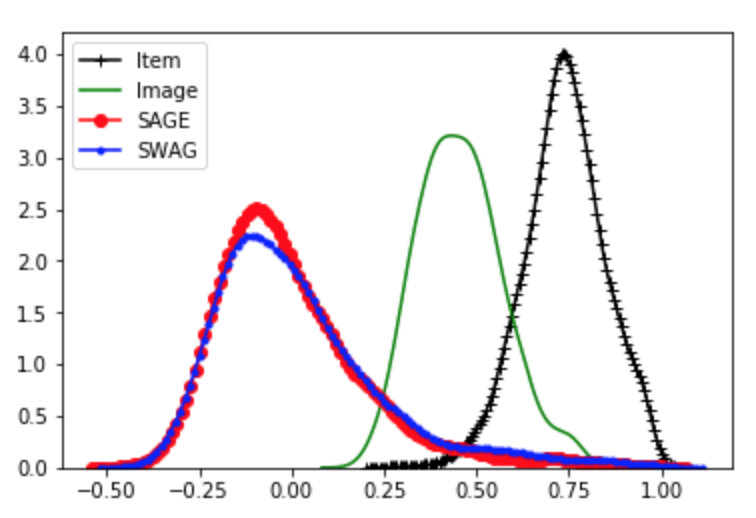}}%
\caption[]{Probability density of pairwise cosine similarity
for image embeddings, text embeddings, SAGE and SWAG embeddings.}%
\label{cos_dist}%
\end{figure*}
\subsection{Impacts of size of neighborhood sampling}
Figure~\ref{vr-time-bench}(a) explains the changes in view-rate as we increase the sample size. In our graphs, each node has more than 100 neighbors. Sampling them leads to homogeneity and also speeds up computation of embeddings in each epoch. For clothing, as evident in the figure, the increase in view rate is marginal beyond sample size of 30. Figure~\ref{vr-time-bench}(b) shows the computational time required for training the model for different sample sizes. It can be seen that the computational time increases significantly beyond sample size of 30. Similar behavior was evident for home category where the graph size was large. Thus, we choose a sample size of $30$ for clothing and home. Electronics and Baby categories have smaller graphs and hence a optimal tradeoff was chosen around sample size of 50. 

\begin{table*}[htb]
\centering
\begin{tabular}{|l|l|l|l|l|l|l|l|l|l|l|}
\hline
View Rate &ID & II & SAGE  & SWAG & SAGE (+ID) & SWAG (+ID) &  SAGE (+II) & SWAG (+II) &  SAGE (+II+ID) & SWAG (+II+ID)  \\ \hline
Clothing   & 16.2 & 10.0   & 10.5 & 10.5 & 22.4                     & \textbf{23.5}                    & 16.5 & 20.2   & 22.5 & \textbf{23.6}                     \\ \hline
Home       & 12.0 & 12.5      & 5.3 & 5.3 & 14.2                & \textbf{16.5}                   & 13.2       & 14.5     & 14.3 & \textbf{16.5}            \\ \hline
Electronic & 20.5 &  20.2       & 7.2 & 7.2 & 21.9                     & \textbf{25.1}                 & 20.5       & 21.5       & 22.1 & \textbf{25.2}           \\ \hline
Baby       & 12.5 &  13.5    & 3.4   & 3.4 & 14                     & 14.5                   & 16.8      & \textbf{17.5}       & 17.0 & \textbf{17.6}           \\ \hline
\end{tabular}
\caption{View rate for different models. ID: Item Description based word2vec embeddings directly used to generate recommendations. II: Item Image based visual embeddings based recommendations. SWAG: SWAG algorithm without any node embeddings, SAGE: GraphSAGE without any node embeddings. $+$ indicates the node embeddings used as input with SAGE and/or SWAG model .}
\vspace{-0.1in}
\label{text-res}
\end{table*}

\subsection{Impacts of aggregators}
All the aggregators presented in this section are weighted i.e. the output of the aggregators is weighted by the edge weight scaled exponentiated by hyperparameter $\gamma$.
The  performance of gcn, swag\_mean, LSTM, mean pooling, maximum pooling are compared in Figure~\ref{agg-func}. The swag\_mean aggregator is same as graphsage\_mean~\cite{hamilton2017inductive} but with weights ($\gamma$). A detailed explanation of these aggregators is given in \cite{hamilton2017inductive}
We find that the swag\_mean and mean\_pooling aggregators outperforms other aggregators by a narrow margin in each category.

\subsection{Impact of input node embeddings}
Table~\ref{text-res} gives view rate of different embeddings for the four categories. ID refers to Item Description based word2vec embeddings directly used to generate recommendations. II refers to Item Image based visual embeddings based recommendations. The numbers for SWAG and GraphSAGE are reported with/without the node embeddings used as input. We make some interesting observations from these view rates: First, we observe that item description embeddings perform slightly better than image embeddings for clothing and almost equal to image embeddings in other categories. This can be attributed to richness of attribute data as well as imperfections in using direct product images for generating embeddings. The product attributes include important information describing the product and are quite useful. The product images have background colors as well as models wearing an outfit. In future iterations, we plan to segment out the targeted product and use embeddings for regions of attention instead of using full image embeddings. We also observed that the SAGE and SWAG models have same performance in absence of node embeddings. The computational time required for SWAG(+ID) is significantly lesser than the time required for SWAG(+II) and SWAG(+ID+II) variants. However, we observe that the performance (view rate) is better than or similar to those. For Baby category, the basic SWAG model has very poor performance but incorporating node embeddings improve the view rates significantly. 

\begin{table*}[h]
\centering
\begin{tabular}{|l|l|l|l|l|l|l|l|l|l|l|}
\hline
Categories & Clothing-5 & Clothing-25 & Home-5  & Home-25 & Baby-5 & Baby-25 &  Elec-5 & Elec-25 \\ \hline
\textbf{MRR (Swag)}   & \textbf{0.083} & \textbf{0.105}  & \textbf{0.061} & \textbf{0.077}  & \textbf{0.062} & \textbf{0.095} & \textbf{0.081} & \textbf{0.088} \\ \hline
MRR (Sage) & 0.077 & 0.085 & 0.044 & 0.052 & 0.048 & 0.053 & 0.060 & 0.068 \\ \hline
MRR (CF) & 0.075 & 0.085 & 0.051 & 0.050 & 0.050 & 0.055 & 0.061 & 0.065 \\ \hline
 \textbf{MPR (Swag)}     & \textbf{0.107} & \textbf{0.146}  & \textbf{0.075} & \textbf{0.098} & \textbf{0.072} & \textbf{0.101} & \textbf{0.106}  & \textbf{0.110} \\ \hline
MPR (Sage) & 0.090 & 0.093 & 0.060 & 0.071 & 0.051 & 0.068 & 0.079 & 0.090 \\ \hline
MPR (CF) & 0.088 & 0.091 & 0.061 & 0.070 & 0.051 & 0.070 & 0.077 & 0.091 \\ \hline
\end{tabular}
\caption{Values of mean percentile ranking (MPR) and mean reciprocal ranking (MRR) in four categories with 5 or 25 recommended items i.e., Clothing-5 indicates the clothing category with top 5 recommended items. We compare the metric values for SWAG with SAGE and collaborative filtering (CF) algorithms.}
\vspace{-0.1in}
\label{mpr-mrr}
\end{table*}

\subsection{MPR and MRR values}
Table~\ref{mpr-mrr} gives the mean Percentile ranking (\textbf{MPR}) and Mean reciprocal ranking (\textbf{MRR})~\cite{Hu2008} metric values for the four categories (Clothing, home, baby and electronics) with 5 or 25 recommended items. We use a session of 2 month customers' real website browsing transactions as a ground truth for our recommendation to calculate the metrics. In Table~\ref{mpr-mrr}, we observe that the SWAG model is uniformly better than SAGE model due to the fact that SWAG model integrates the transactional view information on the edge weight through aggregation. Meanwhile, SAGE and the collaborative filtering recommendation model are almost on the same level in terms of MPR and MRR values. The performance of other baselines presented in literature such as GCN and node2vec was inferior to GraphSAGE  in offline tests, so we chose to not run exhaustive tests on them.

\subsection{Embedding similarity distribution}

An important indication of the effectiveness of the learned embeddings is the widely distributed distances between random pairs of output embeddings. If all items are at about the same distance (i.e., the distances are tightly clustered) then the embedding space does not have enough “resolution” to distinguish between items of different relevance. 
Figure \ref{cos_dist} plots the distribution of cosine similarities between pairs of items
using Image, Item, SAGE and SWAG embeddings. SWAG has the most spread out distribution indicating the ability to distinguish between items of different relevance and also avoiding any collusion in approximate algorithms to find K nearest neighbors (such as LSH).

\subsection{A/B tests}

Lastly, we also report on the production A/B test experiments, which compared the performance of SWAG to other deep learning content-based recommender systems at Target on the task of product recommendations. 

The entire pipeline involves models built in PySpark, TensorFlow and Keras. The input item description or text embeddings are generated using a PySpark word2vec model that runs on a spark cluster sitting on top of HDFS file system. The PySpark modules run on around 150 executors with each driver and executor running on 26G memory. Generation of weights (using Jaccard index and weighted cooccurrences) is done using a PySpark and python map-reduce module respectively. The image embeddings are generated using pre-trained Keras models. The SWAG module is built and run using TensorFlow on GPU nodes cluster with each node having multiple GPUs and connected to HDFS filesystem for convenient access. The TensorFlow version used is 1.10.0. The weights and inputs are refreshed daily and scored by a trained SWAG model to give output embeddings which are used to generate nearest neighbors. To keep the operational latency in the website to be low (few milliseconds), we copy these nearest neighbors to a production server where only a lookup is required in real-time. The recommendations are generated daily to reflect the changes in product catalog and trends in guest browse behaviors. The hyper-parameters for each SWAG model are trained offline and tuning of these parameters, done using skopt library takes a few days for each category. 

The metric of interest is (a) interaction rate or the view rate and (b) conversion rate or the rate of clicked items being finally added to cart and being checked out. These metrics are measured in real-time based on the tagging each and every page and carousel in website. 

We ruled out using raw word2vec based Item-Description embeddings or raw VGG based Item-Image embeddings for production test after testing the recommendations offline with a group of volunteers. The performance of other baselines presented in literature such as GCN and node2vec was inferior to GraphSAGE  in offline tests, so we chose to not test them in A/B tests.
The other deep learning candidates for online test were based on (a) visual recommendations (for clothing) based on CNNs provided by industry vendor, This model uses more complex and multiple CNN models to identify objects of interest in an image and generates similar items such as visual search and shopping tools available publicly online,  (b) visual + behavioral recommendations (for clothing) based on CNNs feeding to a SIAMESE network~\cite{koch2015siamese}, (c) a variant of GRAPHSAGE model with parameters tuned for our graph and item data. We find that SWAG consistently performs better on these metrics than other deep learning based approaches. The performance of (a-c) compared to SWAG as baseline was 50\%, 65\% and 80\% in terms of interaction rate as well as conversion rate.

\begin{figure*}[h]
\centering
\includegraphics[height= 8cm]{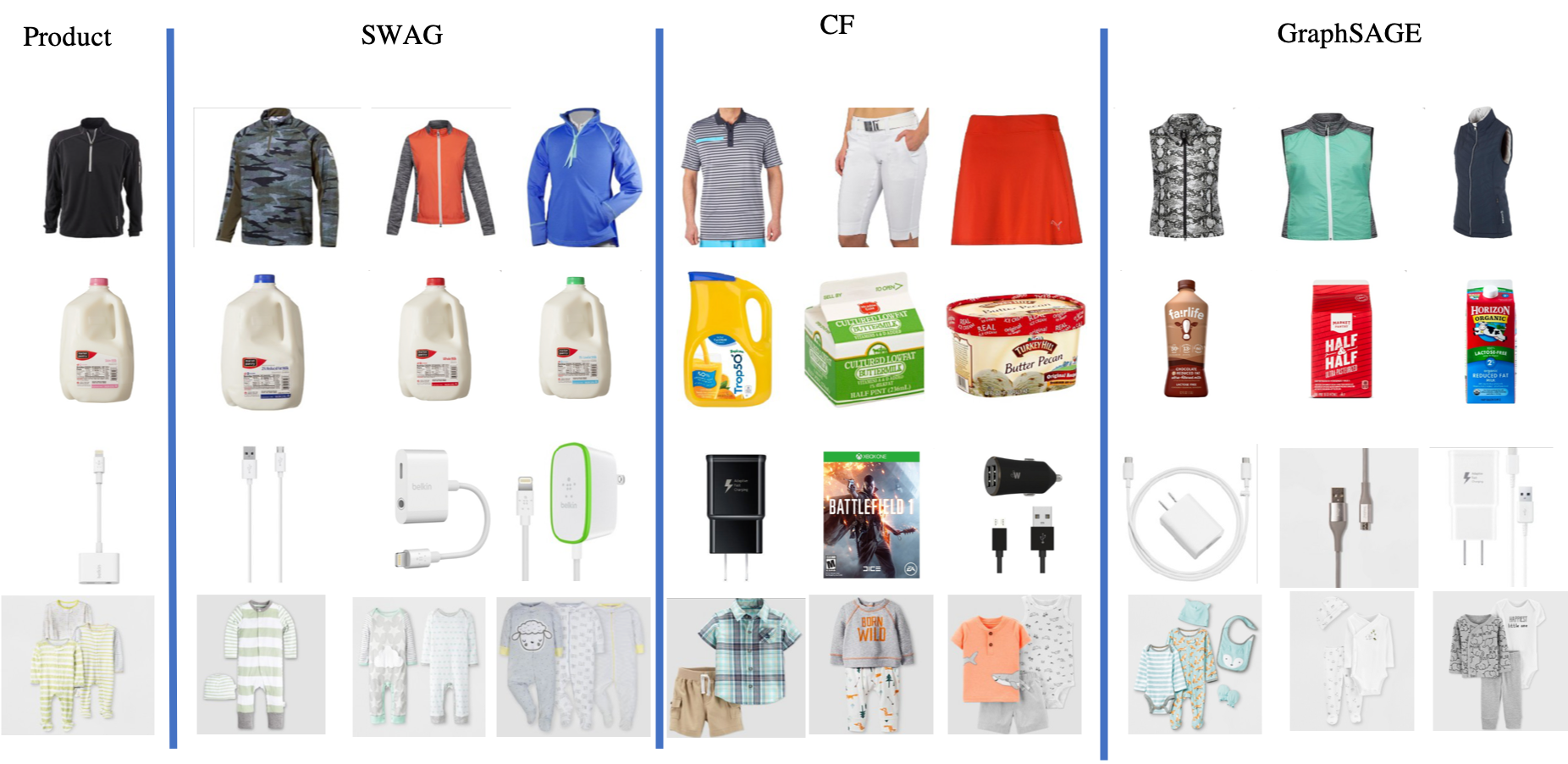}
\caption[Caption for LOF]{Examples of products recommended by different algorithms. Image on extreme left are some random products, while the top 3 recommendations by the raw algorithms (SWAG, SAGE and CF) are presented next}
\label{tcin-rec}
\end{figure*}

\subsection{Visual inspection}

We visualize the embedding space by randomly
choosing 300 items from clothing and compute the 2D t-SNE coordinates in Figure~\ref{img-tsne}. 
The proximity of the SWAG embedding corresponds well with the similarity of content, and that 
items of the same category are embedded closer to each other in t-SNE space. Women's leggings 
are clustered to bottom left while women's athletic tops, shape-wear, socks, men's dresses all form 
their own distant space in t-SNE space. 

Figure \ref{tcin-rec} illustrates top few recommendations using each strategy for 2 sample items. We show the recommendations comparisons of the top three algorithms under consideration (SWAG, graphSAGE and CF) only for a random product from 4 categories - clothing, grocery, accessories and baby. Although all algorithms give good recommendations, the recommendations for SWAG are more nuanced and more similar. For grocery, the browse data is smaller, so we find that CF doesn't do well as others.  CF does tend to give results from other categories than the selected category (juice is recommended for milk, skirt for a fleece and a DVD for an accessory). This is not the case with GraphSAGE and SWAG. However, we find that the top-3 recommendations produced by SWAG are more relevant than GraphSAGE (showing half and half for milk, sleeveless recs for with-sleeve fleece). 

\section{Conclusion}
We proposed SWAG, a  graph convolutional network (GCN) suitable for weighted graphs. SWAG is  capable of learning embeddings for nodes in web-scale graphs and deployed to generate recommendations for millions of product recommendations at Target. We compared performance of SWAG with offline session metrics, embedding distributions, visual  and A/B tests all demonstrating substantial improvements in recommendation performance over other deep learning architectures.

There are possible areas of improvements such as incorporating attention to weighted graphs~\cite{abu2018nips}, more exhaustive evaluation of computer vision models to extract better image embeddings~\cite{huang2017densely} and generating graphs based on store and online purchases made by guests. The authors would also like to demonstrate performance on publicly available weighted datasets (and make some datasets public) in future.

\subsection*{Acknowledgements}
The authors thank Sayon Majumdar and Jacob Portnoy for helpful insights on training and testing SWAG for product recommendations and online tests.

\bibliography{ref}

\begin{thebibliography}{10}

\bibitem{abu2018nips}
Sami Abu-El-Haija, Bryan Perozzi, Rami Al-Rfou, and Alex Alemi.
\newblock Watch your step: Learning graph embeddings through attention.
\newblock In {\em Advances in neural information processing systems}, 2018.

\bibitem{alom2018history}
Md~Zahangir Alom, Tarek~M Taha, Christopher Yakopcic, Stefan Westberg, Mahmudul
  Hasan, Brian~C Van~Esesn, Abdul A~S Awwal, and Vijayan~K Asari.
\newblock The history began from alexnet: A comprehensive survey on deep
  learning approaches.
\newblock {\em arXiv preprint arXiv:1803.01164}, 11 2018.

\bibitem{arora2016simple}
Sanjeev Arora, Yingyu Liang, and Tengyu Ma.
\newblock A simple but tough-to-beat baseline for sentence embeddings.
\newblock 2016.

\bibitem{atwood2016diffusion}
James Atwood and Don Towsley.
\newblock Diffusion-convolutional neural networks.
\newblock In {\em Advances in Neural Information Processing Systems}, pages
  1993--2001, 2016.

\bibitem{bronstein2017geometric}
Michael~M Bronstein, Joan Bruna, Yann LeCun, Arthur Szlam, and Pierre
  Vandergheynst.
\newblock Geometric deep learning: going beyond euclidean data.
\newblock {\em IEEE Signal Processing Magazine}, 34(4):18--42, 2017.

\bibitem{bruna2013spectral}
Joan Bruna, Wojciech Zaremba, Arthur Szlam, and Yann LeCun.
\newblock Spectral networks and locally connected networks on graphs.
\newblock {\em International Conference on Learning Representations (ICLR),
  CBLS}, 2014.

\bibitem{chung1997spectral}
Fan~RK Chung and Fan~Chung Graham.
\newblock {\em Spectral graph theory}.
\newblock Number~92. American Mathematical Soc., 1997.

\bibitem{defferrard2016convolutional}
Micha{\"e}l Defferrard, Xavier Bresson, and Pierre Vandergheynst.
\newblock Convolutional neural networks on graphs with fast localized spectral
  filtering.
\newblock In {\em Advances in Neural Information Processing Systems}, pages
  3844--3852, 2016.

\bibitem{goyal2018graph}
Palash Goyal and Emilio Ferrara.
\newblock Graph embedding techniques, applications, and performance: A survey.
\newblock {\em Knowledge-Based Systems}, 151:78--94, 2018.

\bibitem{grover2016node2vec}
Aditya Grover and Jure Leskovec.
\newblock node2vec: Scalable feature learning for networks.
\newblock In {\em Proceedings of the 22nd ACM SIGKDD international conference
  on Knowledge discovery and data mining}, pages 855--864. ACM, 2016.

\bibitem{hamilton2017inductive}
Will Hamilton, Zhitao Ying, and Jure Leskovec.
\newblock Inductive representation learning on large graphs.
\newblock In {\em Advances in Neural Information Processing Systems}, pages
  1024--1034, 2017.

\bibitem{hamilton2017representation}
William~L Hamilton, Rex Ying, and Jure Leskovec.
\newblock Representation learning on graphs: Methods and applications.
\newblock {\em IEEE Data Engineering Bulletin}, 2017.

\bibitem{he2016identity}
Kaiming He, Xiangyu Zhang, Shaoqing Ren, and Jian Sun.
\newblock Identity mappings in deep residual networks.
\newblock In {\em European conference on computer vision}, pages 630--645.
  Springer, 2016.

\bibitem{Hu2008}
Yifan Hu, Yehuda Koren, and Chris Volinsky.
\newblock Collaborative filtering for implicit feedback datasets.
\newblock In {\em 8th IEEE International Conference on Data Mining}, 2008.

\bibitem{huang2017densely}
Gao Huang, Zhuang Liu, Laurens Van Der~Maaten, and Kilian~Q Weinberger.
\newblock Densely connected convolutional networks.
\newblock In {\em Proceedings of the IEEE conference on computer vision and
  pattern recognition}, pages 4700--4708, 2017.

\bibitem{kipf2016semi}
Thomas~N Kipf and Max Welling.
\newblock Semi-supervised classification with graph convolutional networks.
\newblock {\em ICLR}, 2016.

\bibitem{kipf2016variational}
Thomas~N Kipf and Max Welling.
\newblock Variational graph auto-encoders.
\newblock {\em NeurIPS Workshop on Bayesian Deep Learning (NeurIPS BDL)}, 2016.

\bibitem{koch2015siamese}
Gregory Koch, Richard Zemel, and Ruslan Salakhutdinov.
\newblock Siamese neural networks for one-shot image recognition.
\newblock In {\em ICML Deep Learning Workshop}, volume~2, 2015.

\bibitem{mikolov2013distributed}
Tomas Mikolov, Ilya Sutskever, Kai Chen, Greg~S Corrado, and Jeff Dean.
\newblock Distributed representations of words and phrases and their
  compositionality.
\newblock In {\em Advances in neural information processing systems}, pages
  3111--3119, 2013.

\bibitem{monti2017geometric}
Federico Monti, Michael Bronstein, and Xavier Bresson.
\newblock Geometric matrix completion with recurrent multi-graph neural
  networks.
\newblock In {\em Advances in Neural Information Processing Systems}, pages
  3697--3707, 2017.

\bibitem{ng2002spectral}
Andrew~Y Ng, Michael~I Jordan, and Yair Weiss.
\newblock On spectral clustering: Analysis and an algorithm.
\newblock In {\em Advances in neural information processing systems}, pages
  849--856, 2002.

\bibitem{perozzi2014deepwalk}
Bryan Perozzi, Rami Al-Rfou, and Steven Skiena.
\newblock Deepwalk: Online learning of social representations.
\newblock In {\em Proceedings of the 20th ACM SIGKDD international conference
  on Knowledge discovery and data mining}, pages 701--710. ACM, 2014.

\bibitem{schlichtkrull2018modeling}
Michael Schlichtkrull, Thomas~N Kipf, Peter Bloem, Rianne van~den Berg, Ivan
  Titov, and Max Welling.
\newblock Modeling relational data with graph convolutional networks.
\newblock In {\em European Semantic Web Conference}, pages 593--607. Springer,
  2018.

\bibitem{simonyan2014very}
Karen Simonyan and Andrew Zisserman.
\newblock Very deep convolutional networks for large-scale image recognition.
\newblock {\em ICLR}, 2015.

\bibitem{su2009survey}
Xiaoyuan Su and Taghi~M Khoshgoftaar.
\newblock A survey of collaborative filtering techniques.
\newblock {\em Advances in artificial intelligence}, 2009, 2009.

\bibitem{tang2015line}
Jian Tang, Meng Qu, Mingzhe Wang, Ming Zhang, Jun Yan, and Qiaozhu Mei.
\newblock Line: Large-scale information network embedding.
\newblock In {\em Proceedings of the 24th International Conference on World
  Wide Web}, pages 1067--1077. International World Wide Web Conferences
  Steering Committee, 2015.

\bibitem{ying2018graph}
Rex Ying, Ruining He, Kaifeng Chen, Pong Eksombatchai, William~L Hamilton, and
  Jure Leskovec.
\newblock Graph convolutional neural networks for web-scale recommender
  systems.
\newblock In {\em 24TH ACM SIGKDD Conference on Knowledge Discovery and Data
  Mining}, 2018.

\bibitem{zhou2018graph}
Jie Zhou, Ganqu Cui, Zhengyan Zhang, Cheng Yang, Zhiyuan Liu, and Maosong Sun.
\newblock Graph neural networks: A review of methods and applications.
\newblock {\em arXiv preprint arXiv:1812.08434}, 2018.

\bibitem{zitnik2018modeling}
Marinka Zitnik, Monica Agrawal, and Jure Leskovec.
\newblock Modeling polypharmacy side effects with graph convolutional networks.
\newblock {\em Bioinformatics}, pages 457--466, 2018.

\end{thebibliography}
\bibliographystyle{plain}
\end{document}